**A fiber-integrated single photon source emitting at telecom wavelengths**

Chang-Min Lee[1], Mustafa Atabey Buyukkaya[1], Shahriar Aghaeimeibodi[1], Christopher J. K. Richardson[2], and Edo Waks[1,3,*]

[1]Department of Electrical and Computer Engineering and Institute for Research in Electronics and Applied Physics, University of Maryland, College Park, Maryland 20742, United States

[2]Laboratory for Physical Sciences, University of Maryland, College Park, Maryland 20740, United States

[3]Joint Quantum Institute, University of Maryland and the National Institute of Standards and Technology, College Park, Maryland 20742, United States

**Abstract**

Fiber-coupled single photon sources are essential components of photonics-based quantum information processors. Most fiber-coupled single photon sources require careful alignment between fibers and quantum emitters. In this work, we present an alignment-free fiber-integrated single photon source based on an InAs/InP quantum dot emitting at telecom wavelengths. We designed a nanobeam containing the quantum dots attached to a fiber taper. The adiabatic tapered coupler of the nanobeam enables efficient light coupling to the fiber taper. Using a tungsten probe in a focused ion beam system, we transferred the nanobeam to the fiber taper. The observed fiber-coupled single photon emission occurs with a brightness of 1.5% and purity of 86%. This device provides a building block for fiber-optic quantum circuits that have various applications, such as quantum communication and distributed quantum computing.

Single photons are ideal carriers of quantum information because they can propagate over long distances in optical fibers with extremely low loss[1]. But applications such as quantum communication[2] and photonic quantum computing[3] require high coupling efficiency to the optical fiber mode in order to ensure that the quantum signal faithfully transmits to the receiver or the detector. Most single photon sources emit into free space[4-19] and coupling these sources to fibers necessitates bulky optics that require extremely precise optical alignment. Additionally, this coupling approach is lossy due to imperfect mode-matching. Single photon sources equipped with an efficient coupling scheme into optical fibers could alleviate these problems and would thus be easier to use and integrate with other photonic devices, such as modulators and phase shifters.

Several previous works have reported fiber-coupled single photon sources using either semiconductor quantum dots[20-28] or nitrogen-vacancy centers in diamond[29-33]. The majority of these works control the position of a fiber taper to contact a photonic waveguide on a chip. However, these systems require constant re-alignment and are sensitive to vibrations and temperature fluctuations because the fiber taper and the photonic waveguide move independently. Other works have directly attached the single photon emitters to a fiber taper or a cleaved facet of the fiber[27-32]. However, some of them do not have engineered structures that enable effective mode matching of the single photons into the fiber mode. Besides, these sources emit at wavelengths that are outside the telecom bandwidth where fibers exhibit minimal propagation losses.

In this work, we realize an alignment-free fiber-coupled single photon source at telecom wavelengths. We employed quantum dots emitting near 1300 nm as single photon sources. To guide the single photon emission into a well-defined mode, we fabricated a single-mode nanobeam with a photonic crystal mirror around the quantum dots. Using a tungsten probe installed in a focused ion beam system, we transferred the photonic crystal nanobeam to a tapered optical fiber. The fabricated device achieved a brightness of 1.5% and a single photon purity of 86%. Since we integrate the single photon source into the fiber taper, we can perform pumping, collection, and detection within the optical fiber system, which alleviates the need for complex optical alignment and minimizes the fluctuation of coupling efficiency from mechanical vibrations.

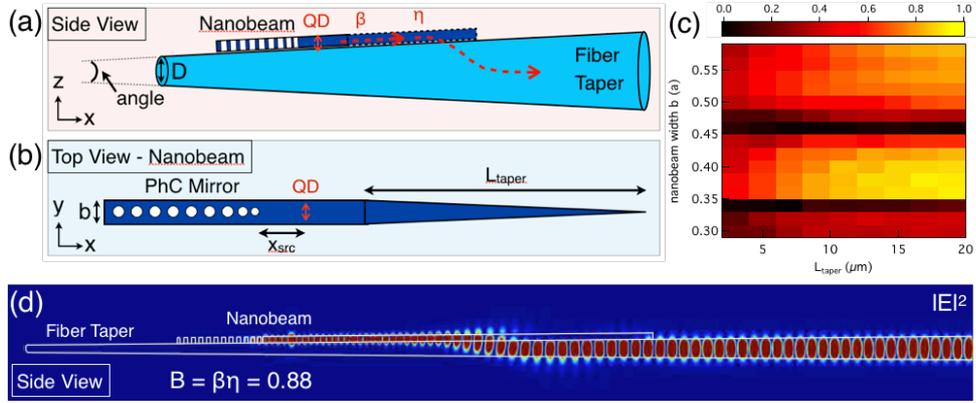

**Figure 1. (a)** Schematic of the integrated photonic crystal nanobeam and the fiber taper. **(b)** Schematic of the photonic crystal nanobeam (top view). b is the width of the nanobeam, PhC is the photonic crystal, QD is a quantum dot, $x_{src}$ is the length between the PhC mirror and the QD, and $L_{taper}$ is the length of the tapered nanobeam. **(c)** A brightness map of the fiber-coupled system as we changed the values of b and $L_{taper}$. **(d)** Electric field intensity map from the side view of the nanobeam/fiber taper construct, with b = 400 nm and $L_{taper}$ = 15 μm. The calculated brightness was 88%.

Figure 1(a) depicts a schematic of the fiber-integrated single photon source. The structure consists of a photonic crystal nanobeam, which contains the InAs/InP quantum dots in the middle (Figure 1(b)), attached on top of a tapered fiber. A photonic crystal mirror composed of an array of etched holes on one end of the beam directs the quantum dot emission in one direction. We smoothly taper one end of the nanobeam to adiabatically transfer the nanobeam-guided single photons to the underlying fiber taper. The gradual fiber taper then transforms the optical mode of the photon into the mode of the bare optical fiber.

An important figure of merit for the device's structure is the brightness, defined as the ratio of the number of single photons collected into the fiber to the number of pump pulses. The brightness is given by the equation $B = q\beta\eta$, where $q$ is the quantum efficiency of the quantum dot, $\beta$ is the single mode coupling efficiency to the nanobeam mode, and $\eta$ is the coupling efficiency to the fiber mode. We optimized the brightness of the designed structure using finite-difference time-domain simulation (Supplementary Note 1). The single photon source has several design parameters: the lattice constant $a$, hole radius $r$ of the photonic crystal mirror, nanobeam thickness $t$, width $b$, taper length $L_{\text{taper}}$, fiber taper tip diameter $D$, and taper angle $\theta_{\text{taper}}$. Figure 1(c) shows the brightness as a function of two parameters: the width of the nanobeam ($b$) and the length of the taper ($L_{\text{taper}}$) with all other parameters fixed (Supplementary Note 2 provides the values used for the remaining parameters). Since the quantum efficiency $q$ is usually close to unity for epitaxially grown quantum dots[34], we assumed $q$ to be unity. The brightness of the device reaches up to 90% for $L_{\text{taper}} \geq 15$ μm and $b \sim 380$ nm (Supplementary Note 3 explains why the brightness dips at $b = 340$ nm and $b = 460$ nm). The brightness of the single photons increases as $L_{\text{taper}}$ gets longer because the smooth taper of the nanobeam improves the adiabatic mode transfer. From the optimization, we selected $b = 400$ nm and $L_{\text{taper}} = 15$ μm. Other design parameters for the fabrication include a lattice constant of $a = 350$ nm, a hole radius of $r = 98$ nm, a thickness of $t = 280$ nm, a fiber taper diameter of $D = 500$ nm, and a fiber taper angle of 1° (Supplementary Note 2). Figure 1(d) shows the electric field intensity profile from the side view, which depicts the smooth adiabatic coupling between the nanobeam and the fiber taper. From the simulation of Figure 1(d), we achieved a fiber-coupled brightness of 88% at $b$ = 400 nm and $L_{taper}$ = 15 μm.

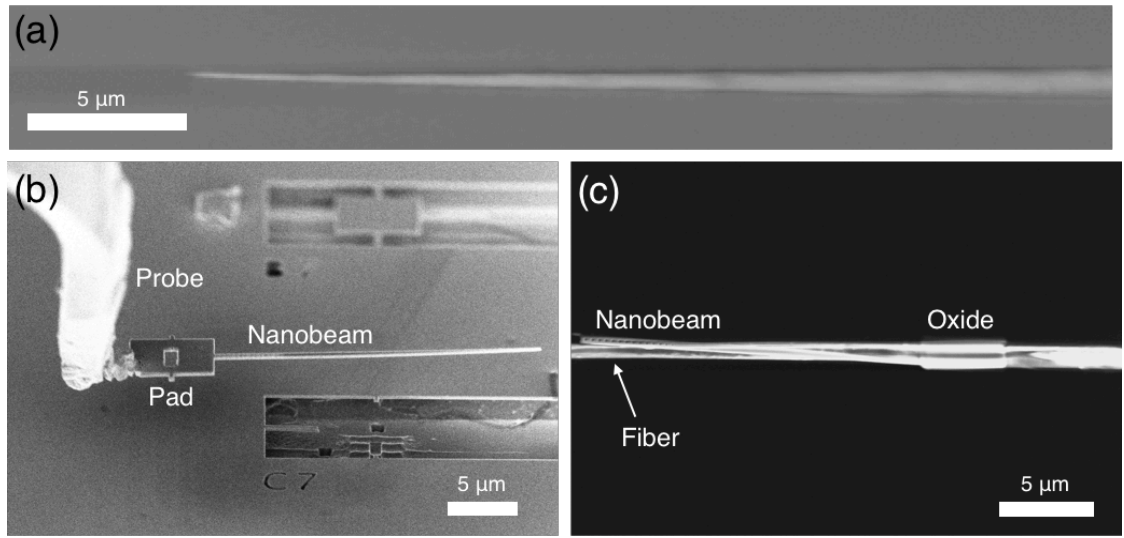

**Figure 2.** Scanning electron micrographs of (a) the fiber taper, (b) the nanobeam being picked up and placed by the tungsten probe, and (c) the fiber-integrated nanobeam. We deposited silicon oxide as a glue between the nanobeam and the fiber taper.

To fabricate the designed structure, we used electron beam lithography and wet chemical etching to pattern an initial wafer containing InAs/InP quantum dots that were grown by molecular beam epitaxy. The wafer consisted of a 280 nm thick InP slab that featured an InAs quantum dot layer at the center, grown on top of a 2 μm thick AlInAs sacrificial layer. We deposited a 220 nm silicon nitride film as an etching mask using plasma-enhanced chemical vapor deposition, followed by electron beam lithography and fluorine-based reactive ion etching to produce the nanobeam pattern on the silicon nitride. Chlorine-based reactive ion etching transferred this pattern onto the InP layer. Then, we removed the sacrificial layer using selective wet etching to make the freestanding nanobeam. We added a rectangular pad a few microns in size on the non-tapered side of the nanobeam to help the transfer process.

We fabricated the fiber tapers from single mode optical fibers using dynamic chemical etching[35]. In this process, we stripped one end of the optical fiber from its coating and dipped it into a 50% hydrofluoric acid (HF) solution for approximately 45 minutes. As the fiber tip etched away, we slowly pulled the fiber out of the solution using a motorized stage. We were able to control the taper angle depending on the HF concentration and the pull-out speed. In this manner, we fabricated fiber tapers with a taper angle of 1° and tip diameter of approximately 100 nm (Figure 2(a)).

We transferred the nanobeam onto the fiber taper using a tungsten probe installed in a focused ion beam system[36]. After placing the probe onto the pad of the nanobeam, we welded the probe tip to the device by depositing silicon dioxide. Then, the ion beam removed the remaining materials tethering the nanobeam to the substrate, allowing us to pick up the structure from the chip (Figure 2(b)) and place it onto the desired region of the fiber taper. In order to ensure attachment, we deposited additional silicon dioxide to "glue" the nanobeam and fiber together. Then, we disconnected the pad from the nanobeam by further ion beam etching. Figure 2(c) shows a nanobeam transferred onto a tapered fiber. As a final step, to improve the mechanical stability of the setup, we coated the entire system with a 50 nm thick aluminum oxide layer using atomic layer deposition.

In order to characterize the properties of the fabricated structure we constructed an all-fiber photoluminescence measurement setup (Supplementary Figure 4). We performed all measurements inside a closed-cycle cryostat that cooled the sample to 4 K. We excited the quantum dots with a Ti:sapphire laser operated at 780 nm in both continuous-wave and pulsed modes with a repetition rate of 76 MHz. In order to measure the spectrum, we spectrally filtered the single photons collected by the fiber using a monochromator and measured the photons using InGaAs array detectors.

When we observed the single photon count rates, we utilized a tunable fiber filter and measured the photons with superconducting nanowire single photon detectors.

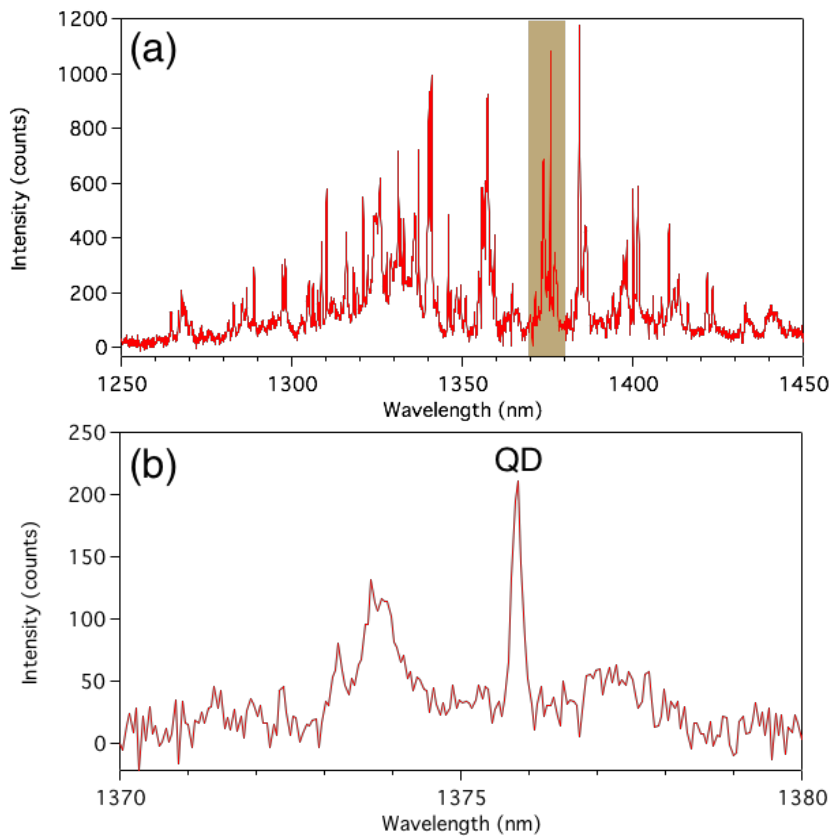

**Figure 3 (a) A photoluminescence spectrum of the fiber-integrated nanobeam produced by a pump power of 3 μW and integration time of 5 s. (b) The magnified photoluminescence spectrum in the orange region of (a) with a pump power of 2.5 μW and integration time of 1 s. We measured the quantum dot peak at 1376 nm.**

Figure 3(a) shows the photoluminescence spectrum obtained by pumping with a 780 nm continuous-wave laser. We observe 20–30 quantum dot lines in the spectrum. Each quantum dot has a different wavelength and position, which creates variations in their coupling efficiency. We isolated a particularly bright and narrow quantum dot line at 1376 nm for detailed analysis. Magnifying the region near 1376 nm (yellow stripe in Figure 3(a)), we observe a narrow peak that corresponds to emission from a single quantum dot (Figure 3(b)).

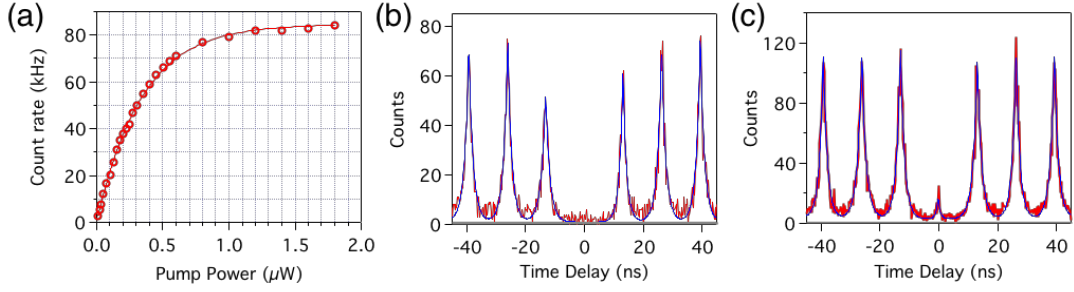

**Figure 4 (a) Pump power dependence of the single photon count rate with pulsed pumping. The red solid line indicates a fitted curve for the calculation of the saturated count rates. (b-c) Second-order autocorrelation measurement with pulsed pumping of (b) 250 nW and (c) 1.2 μW. The blue lines correspond to fitted curves.**

To estimate the fiber-coupled brightness, we measured the single photon count rate as a function of the pump power using the Ti:sapphire laser operated at 780 nm in pulsed mode with a repetition rate of 76 MHz (Figure 4(a)). The pump power dependence of the count rate shows clear saturation behavior. We fit the data to the equation $I(P) = I_0 + I_{max}\left(1 - e^{-\frac{P}{P_{sat}}}\right)$, where $I_0$ is the dark count rate, $I_{max}$ is the maximum emission count rate, and $P$ and $P_{sat}$ are the pump power and the saturation power, respectively. From the fit we determined a maximum count rate of $I_{max}$ = 84 kilo-counts per second (kcps).

To verify that the emission originates from a single quantum dot, we measured second-order autocorrelation. We used a 50/50 fiber beamsplitter to divide the photoluminescence signal into two fibers and measured them with independent single photon detectors. We obtained the coincidence counts between the two detectors as a function of photon arrival time delay using a time-correlated single-photon counter. Figure 4(b) and 4(c) shows the histograms, which represent the second order correlation $g^{(2)}(\tau)$. To calculate the $g^{(2)}(0)$, we normalized the counts at the center within a period (13.16 ns) to the averaged coincidence counts of the nearest three peaks on each side. We also measured the background counts caused by the dark count of the detectors and subtracted them from the coincidence counts[37]. The background counts are relatively small (~1), and we mark them with gray color in Figure 4(b,c) (they can barely be seen). When we set the pump power to 250 nW, which generates a count rate that is half of the saturation rate, $g^{(2)}(0)$ is 0.09 (Figure 4(b)), which shows highly suppressed two-photon emission. When we increased the pump power to 1.2 μW to saturate the quantum dot, $g^{(2)}(0)$ becomes 0.14 due to the background emission by other dots (Figure 4(c))[38].

In order to estimate the fiber-coupled brightness of this system, we measured the transmissions of all the components, such as the fiber vacuum feed-through, fiber filter, couplers, and connectors (Supplementary Table 1). We determined the total system detection efficiency, which includes the transmissions of every component, to be 7%. To calculate the brightness we used the measured single photon count rate $I_{max}$ = 84 kcps, setup detection efficiency T = 7%, the $g^{(2)}(0)$ value of 0.14, and the repetition rate of the pump laser $R$ = 76 MHz. We excluded the additional counts from the multiphoton emission using the $g^{(2)}(0)$ value as $I_{SP} = I_{max} \cdot \sqrt{1 - g^{(2)}(0)}$ = 78 kcps . The collected single photons at the first fiber was $I = \left(I_{max} \cdot \sqrt{1 - g^{(2)}(0)}\right)/T = 1.11$ Mcps . Therefore, the brightness of our fiber-integrated single photon source is $B = I/R = 1.5\%$.

There is a significant discrepancy between the measured and calculated brightness values. In order to understand where the additional loss is coming from, we measured the reflectivity of the fiber-integrated nanobeam with a broadband light source. We found that the amount of back-reflected light was approximately 1%, which means η is approximately 10% (Supplementary Note 4). Thus, we presume that contact and alignment between the nanobeam and the fiber taper are not in the best condition. We also note that the nanobeam sometimes drops from the fiber taper during the cooling procedure (Supplementary Note 5). To prevent the nanobeam dropping, we added a silicon dioxide and aluminum oxide layer after the nanobeam transfer, which possibly causes additional loss.

Our system currently has a brightness of 1.5%, which is similar to that of other alignment-free fiber-coupled quantum dot single photon sources (1–5%)[24,27,28]. However, the current brightness is still far from the theoretical limit of our structure, which is 88%. To improve the brightness of this system, we can promote adhesion between the nanobeam and the fiber taper with either chemical or plasma treatment[39,40]. Besides poor adhesion, there is a possibility that chemical etching of the fiber causes surface roughness, which can give rise to scattering loss. However, it is difficult to characterize the transmission of the etched region of the fiber. As an alternative, we can replace the chemically etched fiber with a flame-pulled fiber taper[41], which can achieve a high transmission (> 90%)[42]. The scattering loss of the flame-pulled fiber can be easily characterized by measuring its transmission, enabling us to investigate whether the main loss is due to the nanobeam-fiber coupling or the fiber taper itself.

In summary, we have demonstrated a fiber-integrated single photon source at telecom wavelengths. We transfer a nanobeam containing InAs/InP quantum dots to a fiber taper using a tungsten probe installed in a focused-ion beam system and obtain a brightness of 1.5% through the fiber. Because of its configuration, this system does not require precise free-space optical alignment or careful positioning of the fiber taper with respect to the quantum dots. We could improve the efficiency by surface treatment of the nanobeam and the fiber taper or by employing a flame-pulled fiber, which could potentially elevate the brightness to as high as 88%. The improved device will pave the way to scalable quantum information processing, such as photonic boson sampling[43] or cluster state generation[44]. Employing nanobeam cavities with the quantum dots, our system could enable fiber-coupled spin-photon interfaces for applications such as quantum phase switches[45,46] or single photon transistors[47].

**Acknowledgements**

The authors would like to acknowledge support from the U.S. Department of Defense, The Center for Distributed Quantum Information at the University of Maryland and Army Research Laboratory, the Physics Frontier Center at the Joint Quantum Institute.

**Supplementary Information**

**A fiber-integrated single photon source emitting at telecom wavelengths**

Chang-Min Lee[1], Mustafa Atabey Buyukkaya[1], Shahriar Aghaeimeibodi[1], Christopher J. K. Richardson[2], and Edo Waks[1,3,*]

[1]Department of Electrical and Computer Engineering and Institute for Research in Electronics and Applied Physics, University of Maryland, College Park, Maryland 20742, United States

[2]Laboratory for Physical Sciences, University of Maryland, College Park, Maryland 20740, United States

[3]Joint Quantum Institute, University of Maryland and the National Institute of Standards and Technology, College Park, Maryland 20742, United States

**Supplementary Note 1 – Numerical Simulation**

We used the three-dimensional finite-difference time-domain method (FDTD solutions, Lumerical) to calculate the brightness and coupling efficiencies. We set the refractive indices of the fiber taper and the InP nanobeam as 1.45 and 3.2, respectively. We calculated the coupling efficiency of the radiation using a point dipole source (a simplified model of quantum dot emission). When we calculated the coupling efficiency, we used a narrow-band Gaussian dipole source (FWHM of 20 nm), while we used a broad-band source for calculating the efficiency spectra (FWHM of 400 nm). We employed non-uniform mesh sizes of 17 nm (~one twentieth of the lattice constant) in a box including the nanobeam, which gets broader outside of the box.

## Supplementary Note 2 – Determining the Structural Parameters

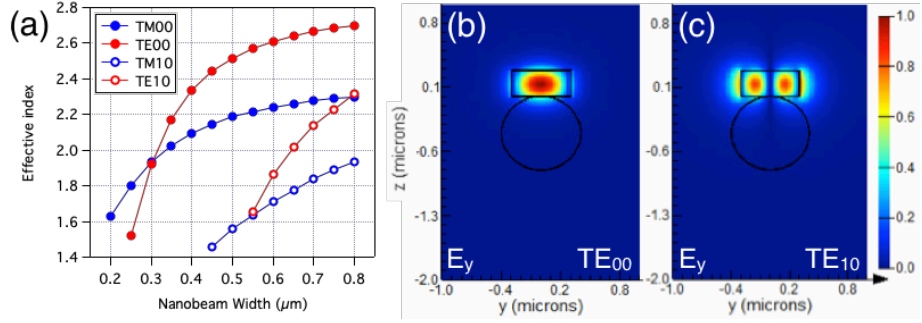

**Supplementary Figure 1.** (a) Effective index of the different modes of the nanobeam on the fiber taper. (b) $E_y$ field profile of the $TE_{00}$ mode. (c) $E_y$ field profile of the $TE_{10}$ mode.

The nanobeam has five structural parameters, including the lattice constant $a$ and hole radius $r$ of the photonic crystal mirror, and the nanobeam thickness $t$, width $b$, and taper length $L_{taper}$. The parameters $a$, $r$, $t$, and $b$ change the bandgap position of the photonic crystal mirror. Additionally, we control $b$ to ensure the single mode condition of the nanobeam, which is approximately 550 nm at a wavelength of 1300 nm (Supplementary Figure 1). We fixed $a$ = 350 nm, $r$ = 98 nm, and $t$ = 280 nm, and adjusted $b$ from 300–580 nm to ensure the bandgap position covered the quantum dot emission wavelength and the nanobeam width met the single mode condition (Supplementary Note 3). We also adjusted the $L_{taper}$ value to obtain adiabatic transfer between the nanobeam and the fiber taper.

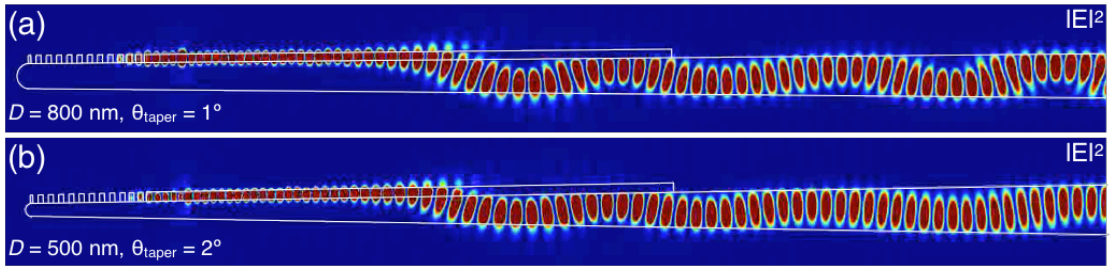

**Supplementary Figure 2.** Electric field intensity map from the side view of the nanobeam/fiber taper construct for (a) D = 800 nm, $θ_{taper}$ = 1° and (b) D = 500 nm, $θ_{taper}$ = 2°. Both calculated field profiles show beating patterns from interference between the fundamental mode and higher-order mode of the fiber taper.

The fiber taper has two structural parameters – the fiber taper tip diameter $D$ and the taper angle $\theta_{taper}$. If $D$ is smaller than ~400 nm, the fiber tip frequently becomes loose and mechanically unstable. In addition, the diameter of the fiber at the mode transfer region should be smaller than the single mode condition (a diameter of 950 nm at 1300 nm wavelength) because the higher-order mode of the fiber taper has a symmetry opposite of the fundamental mode of the untapered fiber and it will eventually leak out. The diameter at the mode transfer region is slightly smaller than the fiber diameter at the right end of the nanobeam ($D_{NBend}$). When $D$ = 500 nm and $\theta_{taper}$ = 1°, ($D_{NBend}$ = 880 nm) the nanobeam mode couples to the fundamental mode of the fiber taper (Figure 1(d) in the main manuscript). When $D$ = 800 nm ($D_{NBend}$ = 1180 nm) the nanobeam mode couples to both the fundamental mode and higher order mode of the fiber taper and shows the beating field profiles (Supplementary Figure 2(a)). Additionally, when we increase the fiber taper angle to 2°, the fiber diameter at the mode transfer region becomes larger than the single mode condition ($D_{NBend}$ = 1270 nm) with $D$ = 500 nm (Supplementary Figure 2(b)). Therefore, we selected a fiber taper tip diameter of 500 nm and a taper angle of 1° to ensure the fiber at the mode transfer region remains in the single mode condition.

**Supplementary Note 3 – Brightness vs. Nanobeam Width**

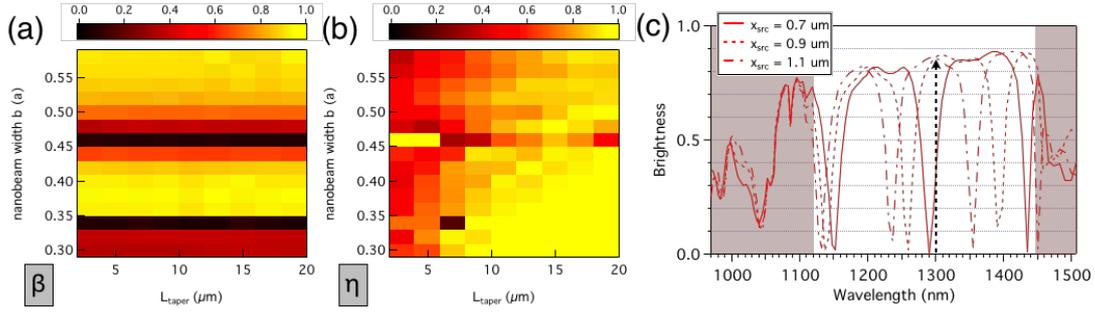

**Supplementary Figure 3.** (a) The single mode coupling efficiency β and (b) fiber coupling efficiency η as a function of $b$ and $L_{taper}$. (c) The brightness spectrum for $b$ = 460 nm and $L_{taper}$ = 15 μm. We also put the dipole source (simulating a quantum dot) at different positions ($x_{src}$) inside the nanobeam and observed the shift of the interference dips.

We calculated the device's brightness as a function of two parameters, the width of the nanobeam $b$ and the nanobeam taper length $L_{taper}$ (Figure 1(c) in the main manuscript). The two-dimensional brightness map features two dips at $b$ = 340 nm and $b$ = 460 nm. From the calculations of the nanobeam mode coupling efficiency $β$ (Supplementary Figure 3(a)) and the fiber coupling efficiency $η$ (Supplementary Figure 3(b)), we confirmed that the dips come from the degraded $β$ factors. In order to investigate the origin of those dips, we calculated the brightness with a broad band Gaussian dipole source at different positions ($x_{src}$) and plotted them as a function of wavelength (Supplementary Figure 3(c)). Since we observed that the dips blueshift as $x_{src}$ increases, we can attribute those dips in the brightness to the interference between the emitted photon from the dipole and the reflected photon at the photonic crystal mirror.

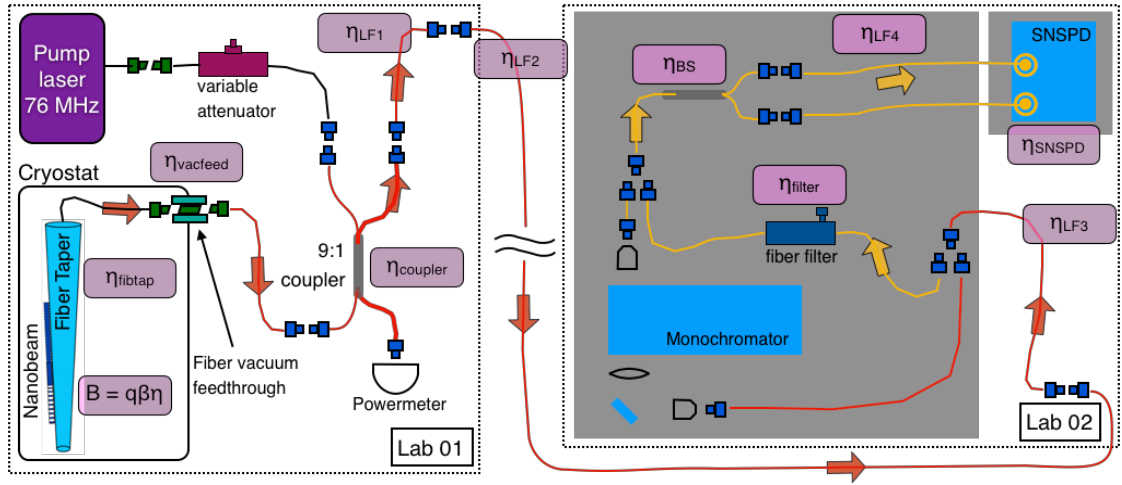

**Supplementary Figure 4.** All-fiber measurement setup demonstrating the flow of the collected single photons. $\eta_{fibsca}$ is the transmission at the fiber taper, $\eta_{vacfeed}$ is the transmission at the fiber-type vacuum feed-through, $\eta_{coupler}$ is the transmission at the fiber coupler, $\eta_{LF1}$, $\eta_{LF2}$, $\eta_{LF3}$, and $\eta_{LF4}$ are the transmissions of the long fibers, $\eta_{filter}$ is the transmission at the fiber-type spectral filter, $\eta_{BS}$ is the transmission at the fiber-type beam splitter, and $\eta_{SNSPD}$ is the detection efficiency of the superconducting nanowire single photon detector. All the transmissions include the fiber connection loss.

**Supplementary Table 1. Transmission of each component.**

| Component | Transmission |
| --- | --- |
| $\eta_{vacfeed}$ | 0.51 |
| $\eta_{coupler}$ | 0.89 |
| $\eta_{LF1}\eta_{LF2}\eta_{LF3}$ | 0.82 |
| $\eta_{filter}$ | 0.29 |
| $\eta_{BS}$ | 0.89 |
| $\eta_{LF4}$ | 0.91 |
| $\eta_{SNSPD}$ | 0.80 |

**Supplementary Note 4 – Reflectivity Measurement**

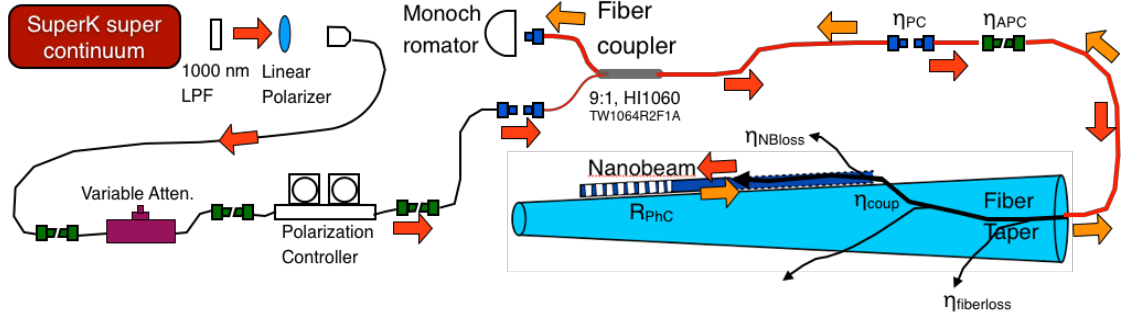

**Supplementary Figure 5.** The reflectivity measurement setup. Red (orange) arrow indicates the direction of the input (back-reflected) light. LPF: long pass filter. $\eta_{PC}$ is the transmission at the FC/PC fiber connector, $\eta_{APC}$ is the transmission at the FC/APC fiber connector, $\eta_{coup}$ is the coupling efficiency between the nanobeam and the fiber taper, $\eta_{NBloss}$ is the transmission at the nanobeam, $\eta_{fiberloss}$ is the transmission at the fiber taper, $R_{PhC}$ is the reflectivity at the photonic crystal mirror.

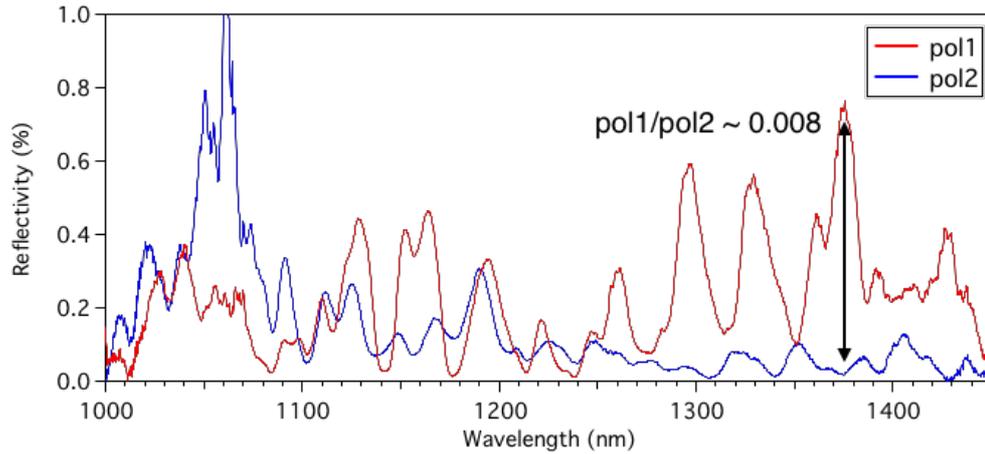

**Supplementary Figure 6.** A reflectivity spectrum with two different input polarizations. pol1 (pol2) is an input polarization that makes the reflected light at ~1376 nm maximum (minimum). We used a super-continuum source as an input light and measured the back-reflected light with a fiber coupler. We divided the back-reflected light spectrum with the spectrum of the input light to obtain the reflectivity.

We performed reflectivity measurements with the all-fiber configuration (Supplementary Figure 5). From the reflectivity spectra of our device (Supplementary Figure 6), we obtained a reflectivity of 0.8% at 1376 nm. When we estimate the reflected light $P_r$ with the transmission of the components,

$$P_r = P_i \eta_{PC} \eta_{APC} \eta_{fiberloss} \eta_{coup} \eta_{NBloss} R_{PhC} \eta_{NBloss} \eta_{coup} \eta_{fiberloss} \eta_{APC} \eta_{PC} \eta_{fibcoupler} \quad (1)$$

$$= (\eta_{PC})^2 (\eta_{APC})^2 (\eta_{fiberloss})^2 (\eta_{NBloss})^2 (\eta_{coup})^2 R_{PhC} \eta_{fibcoupler} P_i. \quad (1)$$

Here, $\eta_{fibcoupler}$ is the transmission at the fiber coupler. Therefore,

$$\eta_{coup} = \sqrt{\frac{P_r}{P_i} \frac{1}{(\eta_{PC})^2 (\eta_{APC})^2 (\eta_{fiberloss})^2 (\eta_{NBloss})^2 (\eta_{coup})^2 R_{PhC} \eta_{fibcoupler}}}. \tag{1}$$

If we set the $\eta_{PC} = 0.95$, $\eta_{APC} = 0.9$, and $\eta_{fibcoupler} = 0.95$, which are measured values, and assume that $\eta_{fiberloss} = \eta_{NBloss} = R_{PhC} = 1$, the coupling efficiency between the nanobeam and the fiber taper $\eta_{coup} = 13.2\%$.

**Supplementary Note 5 – Nanobeam dropping during the cool-down**

When we cool down the fabricated samples for the actual measurement, we monitor the broad photoluminescence signal from room temperature to 4 K. Without the aluminum oxide coating (made with atomic layer deposition) and the silicon dioxide deposition, the nanobeam frequently drops off the fiber taper during the cooling process, typically in the 100–200 K range. This dropping is possibly due to the different thermal expansion coefficient between InP and $SiO_2$. After we start to apply the oxide coating, the probability of the nanobeam dropping significantly reduces (from 70–80% to 20–30%, roughly). Therefore, we infer that the adhesion between the nanobeam and the fiber taper is not very strong.